\documentclass[twocolumn, aps, prb, 10pt,%
	superscriptaddress, nobalancelastpage, raggedbottom]{revtex4-1}

\usepackage{amsmath}
\usepackage{graphicx}

\newcommand{\dvec}[1]{\ensuremath{\boldsymbol{#1}}}
\renewcommand{\vr}{\dvec{\mathrm{r}}}
\newcommand{\vk}{\dvec{\mathrm{k}}}

\newcommand{\dmudn}{\frac{d\mu}{dn}}
\newcommand{\nimp}{n_i}
\newcommand{\nrms}{n_{\mathrm{rms}}}
\newcommand{\nav}{\langle n\rangle}
\newcommand{\muav}{\langle\mu\rangle}
\newcommand{\gu}{\gamma_1^2 + u^2}

\newcommand{\eV}{\mathrm{eV}}
\newcommand{\meV}{\mathrm{meV}}
\newcommand{\cmsq}{\mathrm{cm}^{-2}}

\begin{document}
\title{Inhomogeneity and nonlinear screening in gapped bilayer graphene}
\author{D. S. L. Abergel}
\affiliation{Condensed Matter Theory Center, Department of Physics,
University of Maryland, College Park, Maryland 20742, USA}

\author{E. Rossi}
\affiliation{Department of Physics, College of William and Mary,
Williamsburg, Virginia 23187, USA}

\author{S. Das Sarma}
\affiliation{Condensed Matter Theory Center, Department of Physics,
University of Maryland, College Park, Maryland 20742, USA}

\begin{abstract}
	We demonstrate that for gapped bilayer graphene, the nonlinear
	nature of the screening of an external disorder potential and the
	resulting inhomogeneity of the electron liquid are crucial for
	describing the electronic compressibility.
	In particular, traditional diagrammatic methods of many-body theory
	do not include this inhomogeneity and therefore fail to reproduce
	experimental data accurately, particularly at low carrier densities.
	In contrast, a direct calculation of the charge landscape via a
	numerical Thomas-Fermi energy functional method along with the
	appropriate bulk averaging procedure captures all the essential
	physics, including the interplay between the band gap and the
	inhomogeneity.
\end{abstract}

\maketitle

\section{Introduction}

Measurements of the electronic compressibility provide a way of
characterizing the electron gas in both three-dimensional and
two-dimensional materials and information about the nature of
interactions between the electrons and the influence of the environment
on the electron liquid can be gained. Therefore, it is highly important
to have a clear theoretical understanding of experimental measurements
of the compressibility.
The compressibility $K$ is given by\cite{abergel-ssc152} $K =
\frac{1}{n^2}\frac{dn}{d\mu}$ where $n$ is the excess carrier density
and $\mu$ is the chemical potential and so the key calculation is that
of $\dmudn$.
Recently, the compressibility of both monolayer\cite{dassarma-rmp83} and
bilayer\cite{abergel-advphys59} graphene has been examined with
capacitance probes, \cite{henriksen-prb82, young-arXiv1004, young-prb84}
scanning single electron transistor (SET) microscopy
\cite{martin-natphys4, martin-prl105} and scanning tunneling microscopy
(STM).
\cite{zhang-y-natphys5, deshpande-prb79, xue-natmat10, deshpande-apl95}
In STM, the fine spatial resolution of recent studies has shown a high
degree of inhomogeneity in the charge landscape of graphene systems,
and revealed that the material used as the substrate has a
significant impact on this inhomogeneity.\cite{xue-natmat10}  When the
overall excess electronic density is close to zero (the so-called
``charge-neutrality point''), the electron liquid breaks up into
`puddles' of electrons and holes, presumably to screen an external
potential generated by disorder of some kind.  Theoretical studies of
graphene systems with charged impurities \cite{rossi-prl107} and
corrugations or ripples \cite{gibertini-prb85} have shown that either of
these mechanisms may contribute to the observed inhomogeneity.

Also, gapped electronic systems are highly important in many device
applications, and bilayer graphene is an attractive material in this
context since the band gap and carrier density can be controlled
dynamically via gating. \cite{dassarma-rmp83, abergel-advphys59,
mccann-prl96} 
Therefore a clear understanding of the interplay between the
inhomogeneity which is intrinsic to all graphene systems and the gapped
nature of gated bilayer graphene is essential. In this paper we
present a full analysis of
this issue via the theoretical consideration of the compressibility and
comparison with recent experimental work. \cite{henriksen-prb82}
Consistent with experimental findings (which are described in detail in
Ref. \onlinecite{dassarma-rmp83}), we take the
disorder to be arising from random quenched charged impurities in the
environment of the bilayer graphene (BLG) with a two-dimensional (2D)
impurity density of $\nimp$ separated
from the graphene layers by an average distance $d$.
In Section \ref{sec:perttheo} we apply the standard diagrammatic
perturbation theory which is widely used to describe electron-impurity
scattering in condensed matter systems. 
We show that this theoretical technique does not give the correct
qualitative picture in the gapped regime when the inhomogeneity is
strong.
We contend that this failure is due to the fact that this theory cannot
incorporate the effects of the inhomogeneous charge distribution or the
nonlinear nature of the screening. 
In Sec. \ref{sec:tft} we present a functional approach to calculate
$\dmudn$ based on and extending the Thomas-Fermi theory (TFT) of Ref.
\onlinecite{rossi-prl107}, that is able to take into account the effect
on the compressibility of the interplay of disorder and band-gap in the
theoretically challenging regime when the band gap is of the same order,
or smaller, than the strength of the disorder.

We shall show that there are in fact two different criteria for
assessing when the inhomogeneity is too strong for perturbative theories
to be valid. The first is when the proportion of the graphene which is
in the insulating (and hence incompressible) state becomes significant.
The second is when the average fluctuations characterized by the root
mean square of the density distribution becomes large compared to the
average carrier density.
This situation is qualitatively different in monolayer graphene
\cite{abergel-ssc152, rossi-prl101, asgari-prb77, hwang-prl99} where
there is no band gap and hence the screening nonlinearity has a much
smaller effect on the compressibility since there is no mixed phase.

\section{The clean limit}

\begin{figure}
	\includegraphics[]{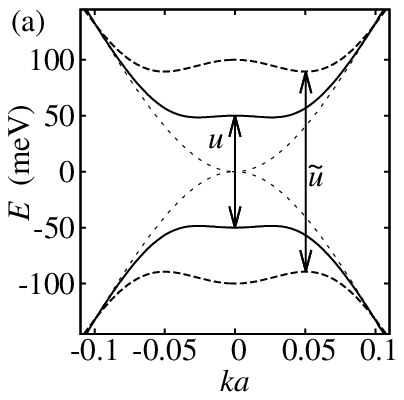}
	\hfill
	\includegraphics[]{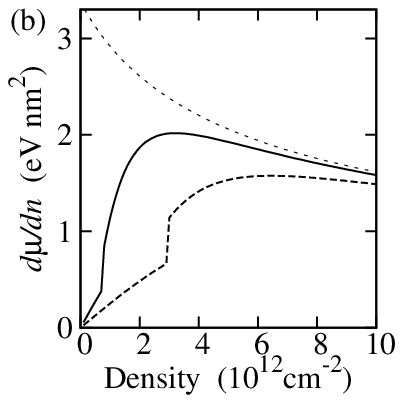}
	\caption{(a) The low-energy conduction and valence bands of bilayer
	graphene with and without a gate-induced band gap. 
	(b) The clean $\dmudn$ given by Eq. \eqref{eq:dmudn}.
	For both panels the dotted lines are $u=0$, the solid lines are
	$u=100\meV$, and the dashed lines are $u=200\meV$. 
	The wave vector is measured in units of the lattice constant $a$.
	\label{fig:bandstructure}}
\end{figure}

In this section we give an overview of the single particle physics of
bilayer graphene in order to remind the reader of the most important
points and to define our notation.
The band structure of bilayer graphene can be approximated via a
four-component Hamiltonian which describes the wave function amplitude
on each of the four lattice sites in the unit cell.
\cite{abergel-advphys59} In this representation, there are two branches
in the conduction and valence bands, with one branch separated from the
other by the inter-layer coupling energy $\gamma_1\approx 0.4\eV$. 
In the strongly
inhomogeneous regime, the split bands may become partially
occupied even at low average carrier density and therefore we keep all
four bands in our analysis. The band structure is given by
\begin{equation}
	E_{\alpha\vk} = \nu_\alpha \sqrt{ v_F^2 k^2 + \frac{\gamma_1^2}{2} +
	\frac{u^2}{4} + b_\alpha \sqrt{ \frac{\gamma_1^4}{4} 
	+ v_F^2 k^2( \gamma_1^2 +u^2 )}}
	\label{eq:bandstructure}
\end{equation}
where $u$ is the band gap at $k=0$, $\nu=+1$ in the conduction band and
$\nu=-1$ in the valence band, $b=+1$ in the split branches and $b=-1$ in
the low-energy branches, $v_F$ is the Fermi velocity associated with
monolayer graphene, and $\hbar=1$. These bands are each four-fold
degenerate due to the presence of spin and the two valleys in the
Brillouin zone.
The actual band gap\cite{mccann-prl96} is given by $\tilde u =
u\gamma_1/\sqrt{u^2+\gamma_1^2}$.
This band structure is illustrated in Fig. \ref{fig:bandstructure}(a)
for the ungapped case and two different band gaps. The quartic (or
`sombrero') shape of the low-energy branches is clearly visible in the
gapped examples. 
\begin{widetext}
The compressibility associated with these bands can be calculated
analytically by relating the density to the Fermi energy and computing
the derivative. \cite{abergel-ssc152}
The full expression is
\begin{equation}
	\frac{d\mu}{dn} = \frac{\gamma_1 u}{\sqrt{\gu}} \delta(n)
	+ \frac{v_F^2\pi}{2}
	\begin{cases}
	\frac{v_F^2\pi |n|}{\sqrt{\gu}} \frac{1}{\sqrt{v_F^4 \pi^2 |n|^2 + u^2
		\gamma_1^2}} & v_F^2 \pi |n| < u^2 \\[0.5cm]
	\frac{1 - \frac{\gu}{2\sqrt{v_F^2 \pi |n|(\gu) + \frac{\gamma_1^4}{4}}}}
		{\sqrt{v_F^2 \pi |n| + \frac{u^2}{4} + \frac{\gamma_1^2}{2}
		- \sqrt{v_F^2 \pi |n|(\gu) + \frac{\gamma_1^4}{4}}}} &
		u^2 \leq v_F^2 \pi |n| < 2\gamma_1^2 + u^2 \\[1.0cm]
	\frac{1}{\sqrt{2v_F^2 \pi |n| - u^2}} & 
		v_F^2 \pi |n| \geq 2\gamma_1^2 +u^2
	\end{cases}
	\label{eq:dmudn}
\end{equation}
where $v_F^2 \pi n$ is the Fermi energy measured in terms of
the density. 
\end{widetext}
Note that the first term implies that the electron liquid is
incompressible at $n=0$. 
There are three different cases because of the
changes in the topology of the Fermi surface. At $n\approx 0$ the Fermi
surface is ring shaped, but when the Fermi energy reaches the top of the
sombrero part of the band structure (i.e. $\mu = u$), this changes to a
disk and hence there is a step in the value of $\dmudn$. Then, at much
higher density ($\sim 2\times 10^{13}\cmsq$) the split band becomes
occupied, and there are now two Fermi surfaces so that there is a second
jump in $\dmudn$.
The clean $\dmudn$ is shown in Fig. \ref{fig:bandstructure}(b) for the
three cases corresponding to the band structures in Fig.
\ref{fig:bandstructure}(a).

\section{Diagrammatic perturbation theory \label{sec:perttheo}}

In this section we discuss perturbative methods for describing the
electron-impurity interaction. 
Inherent in this approach \cite{mahan-1993} is an average over disorder
realizations which explicitely restores translational symmetry to the
theory. We shall show that this approximation is not valid in the
situation we discuss because it removes the possibility for
inhomogeneity to form as a consequence of the electron-impurity
scattering.
In order to show that this failure is not an artifact of the specific
level of approximation in the theory, we also apply the perturbation
expansion keeping two different sets of diagrams, sum the infinite
series associated with them, and find the same qualitative
features in the predicted $\dmudn$ which do not match the experimental
results.

In order to obtain $\dmudn$ in this microscopic theory, the crucial
feature which distinguishes the disordered case from the clean case is
the presence of the electron-impurity self-energy in the electron
Green's function. 
We compute this self-energy within two different approximations
\cite{mahan-1993, doniach-1998} -- the Born approximation (BA) and the
self-consistent Born approximation (SCBA).
In the BA the self-energy is
\begin{equation}
	\Sigma^{\mathrm{BA}}_\alpha(\vk,E) = \nimp \sum_{\vk',\alpha'}
	\frac{|V(\vk-\vk')|^2 F_{\alpha\alpha'}(\vk,\vk')}
	{E - E_{\alpha'\vk'} + i\eta},
\end{equation}
where $F_{\alpha\alpha'}(\vk,\vk')$ is the wave function overlap of the
initial and final states in the scattering process, $E_{\alpha\vk}$ is
the energy of an electron with wave vector $\vk$ in band $\alpha$ from
Eq. \eqref{eq:bandstructure}, $\eta$ is a positive infinitesimal, and
$V(\vk)$ is the screened impurity potential
\begin{equation}
	V(\vk) = \frac{2\pi e^2}{\kappa(k+q_s)} e^{-kd},\quad
	q_s = \frac{2\pi e^2}{\kappa} \rho_0(\mu),
\end{equation}
with $q_s$ being the screening wave vector in the static random
phase approximation, $\rho_0$ the density of states of the clean system, 
and $\kappa$ the dielectric constant. 
Note that the assumption of a homogeneous charge
landscape also enters in the use of $q_s$ as the screening wave vector.
The SCBA takes into account the full Green's function for propagation
between scattering events and for which the self-energy is given by
\begin{equation}
	\Sigma^{\mathrm{SCBA}}_\alpha(\vk,E) = \nimp \sum_{\vk',\alpha'}
	\frac{|V(\vk-\vk')|^2 F_{\alpha\alpha'}(\vk,\vk')}
	{E - E_{\alpha'\vk'} - \Sigma_{\alpha'}(\vk',E)}.
\end{equation}
Note the self-consistent inclusion of the self-energy in the Green's
function on the right-hand side. 
Once the self-energy has been obtained, the electron Green's
function $G_\alpha(\vk,E) = \left[ E - E_{\alpha\vk} -
\Sigma_{\alpha}(\vk,E) \right]^{-1}$ can be straightforwardly computed. 
The density of states $\frac{dn}{d\mu}$ is then related to the imaginary
part of the Green's function, since
\begin{equation}
	\rho(E) \equiv \frac{dn}{d\mu} = - \frac{g_s g_v}{\pi}
	\sum_\alpha \int \frac{d^2 \vk}{4\pi^2} \mathrm{Im} G_\alpha(\vk,E),
\end{equation}
where $g_s = g_v = 2$ are the spin and valley degeneracies,
respectively.

\begin{figure}
	\centering
	\includegraphics[]{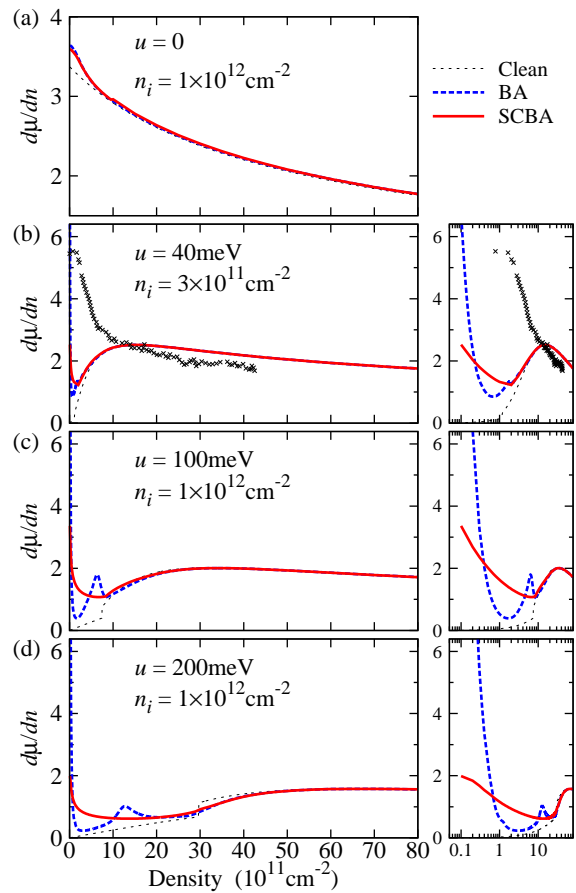}
	\caption{(Color online) $\dmudn$ for three different values of the
	band gap and impurity density for the diagrammatic perturbation
	theories. The points in panel (b) are taken from Fig. 2 of
	Ref. \onlinecite{henriksen-prb82}. Throughout, $\dmudn$ is given in
	units of $\eV \mathrm{nm}^2$.
	\label{fig:perttheo}}
\end{figure}

In Fig. \ref{fig:perttheo} we show the calculated $\dmudn$ for the clean
case, \cite{abergel-ssc152} the BA, and the SCBA as a function of the
carrier density. The right-hand panels show the same data on a
logarithmic scale to emphasize the low-density features. In the ungapped
case shown in Fig. \ref{fig:perttheo}(a), the BA and SCBA give
essentially the same result as the clean limit. When a band gap is
present, as in Figs. \ref{fig:perttheo}(b) to \ref{fig:perttheo}(d), 
the clean limit shows a clear step occuring at the density $\tilde n =
u^2/(v_F^2 \pi)$
which marks the density where the chemical potential leaves the sombrero
region of the band structure and the topology of the Fermi surface
changes from a ring to a disk. \cite{abergel-advphys59} For $n \gg
\tilde{n}$ the BA and SCBA are similar to the clean limit, but for
low-to-moderate density $n \lesssim \tilde n$ the strong modification
of the density of states (DOS) near the band edge \cite{abergel-prb85}
implies that $\dmudn$ is enhanced relative to the clean system but is
still a decreasing
function as $n$ becomes small. There is also a sharp divergence in the
BA and SCBA for very small $n$ which is not observed experimentally.
\cite{henriksen-prb82, young-arXiv1004} Additional structure for
$n<\tilde n$ in the BA comes from the non-trivial shape of the DOS near
$\mu=u$ in that approximation, which is smoothed out by the self
consistency of the SCBA. \cite{abergel-prb85}

In Fig. \ref{fig:perttheo}(b) we also show experimental data for the
gapped regime. Note that the data in Ref. \onlinecite{henriksen-prb82}
is a capacitance measurement, and in order to extract $\dmudn$ we need
accurate knowledge of experimental parameters, such as the impurity
density, the gate-induced band gap, the stray capacitance, and the
dielectric environment, all of which are known only approximately.
Therefore the
experimental data shown here does not correspond to the
parameters used in the calculation and hence we cannot expect
quantitative agreement. Other experiments \cite{young-arXiv1004,
young-prb84} show the same qualitative features as in Ref.
\onlinecite{henriksen-prb82} although direct comparison to these data
is not possible since the low-density $\dmudn$ is obscured by the
specifics of the experimental setup used in these measurements. 
We see in Fig. \ref{fig:perttheo}(b) that a broad peak forms in the
experimental data at low density in complete qualitative contrast to the
BA and SCBA theoretical results.
Therefore, these theories utterly fail to capture the essential physics
of the gapped system at low densities.
This is, however, not unexpected since the low density regime is
completely dominated by the charged impurity induced random puddles of
compressible and incompressible regions.

\section{Thomas-Fermi theory \label{sec:tft}}

We now describe the TFT for the inhomogeneous system. 
In this approach, the carrier density landscape is obtained by
minimizing a Thomas-Fermi energy functional of the spatially-varying
density $n(\vr)$ that includes a term due to the presence of disorder.
The TFT is similar in spirit to the density functional theory (DFT)
\cite{hohenberg-pr136, kohn-pr140, kohn-rmp71} but in TFT the kinetic
energy operator is also replaced by a functional $E_K[n]$. This
simplification makes the TFT valid only when the density profile varies
on length scales larger than the Fermi wavelength, i.e., when $|\nabla
n/n| < k_F$, where $k_F=\sqrt{\pi |n|}$ is the Fermi wave vector. 
This approach has been very successful in the context of transport
calculations \cite{dassarma-rmp83} which provides strong
phenomenological support for the use of this theory.
Specifically for bilayer graphene, the puddle length scale is $\sim
20\mathrm{nm}$ and the density of carriers in the puddles is $\sim
10^{12}\cmsq$ so that this inequality is marginally satisfied. 
However, we can also justify this approximation by pointing out
that the root-mean-square of the density distribution is much larger
than its average. 
At the charge-neutrality point (CNP) the average density $n=0$
cannot be taken as a measure of the typical carrier density inside the
puddles and a better estimate is given by $\nrms$. As a consequence, at
low dopings (close to the CNP) $\nrms$ should be used instead of
$n$ in the inequality above. 
Given that $\nrms \sim \nimp$ we then conclude that the TFT is
valid at all densities so long as $\nimp$ is not too small ($\nimp >
10^{11}\cmsq$).
A full DFT for the disordered problem has been completed for monolayer
graphene \cite{polini-prb78} and shows very similar results to the TFT
applied in the same context. \cite{rossi-prl101}
However, the DFT is much more demanding of computational resources, and
therefore it is not possible to simulate large lattice sizes or complete
a comprehensive average over disorder realizations in a reasonable
timescale. 
Moreover, given the difficulty to quantitatively compare the theoretical
and experimental results (since parameters, such as the impurity density
and stray capacitance, are not known accurately) and the complete failure
of the diagrammatic methods to even achieve a gross qualitative
description of the experimental results at low carrier density, our main
motivation is to show that a functional method like the TFT is able to
capture the qualitative features of the compressibility observed in
experiments. 
The functional approach described below, notwithstanding the specifics
of the TFT, is more than adequate to describe the compressibility of gapped
systems in which the band gap is comparable to or smaller than the
disorder strength.

The TFT energy functional is given by
\begin{multline}
	E[n] = 
	E_K[n(\vr)] 
	+ \frac{e^2}{2\kappa}\int d\vr' \int d\vr 
	\frac{n(\vr)n(\vr')}{|\vr-\vr'|} \\
	+ \frac{e^2}{\kappa} \int d\vr V_D(\vr)n(\vr) - \mu \int d\vr
	n(\vr),
	\label{eq:Efunc}
\end{multline}
where $e^2 V_D/\kappa$ is the bare disorder potential which is assumed
to be due to the Coulomb interaction with random charged impurities with
no spatial correlation and an equal probability of being positively or
negatively charged.
\begin{widetext}
The first term is the kinetic energy where
\begin{equation}
	\epsilon_K[n(\vr)]
	\equiv \frac{\delta E_K}{\delta n}
	= \begin{cases}
	\frac{1}{2} \sqrt{\frac{ v_F^4 \pi^2 |n|^2 + \gamma_1^2 u^2}{ \gu }},
		& v_F^2\pi |n| < u^2 \\
	\sqrt{v_F^2\pi |n| + \frac{\gamma_1^2}{2} + \frac{u^2}{4}
		-\sqrt{ v_F^2 \pi |n|(\gu) + \frac{\gamma_1^4}{4}} },
		& u^2 \leq v_F^2\pi |n| < 2\gamma_1^2 + u^2 \\
	\frac{1}{2} \sqrt{ 2 v_F^2 \pi |n| - u^2}, & 2\gamma_1^2 + u^2 \leq
		v_F^2 \pi |n|
	\end{cases}
\end{equation}
is the ground state kinetic energy per excess carrier.
\end{widetext}
The second term is the Hartree part of the
electron-electron
interaction, the third term is the contribution due to the disorder
potential, and in the fourth term $\mu$ is the chemical potential. We
neglect exchange and correlation terms \cite{borghi-prb82} since, as we
shall show below, $\dmudn$ at low density is predominantly determined 
by the proportion of the sample which is incompressible and the
inclusion of these terms will not change this. The ground state density
landscape is identified by the equation $\delta E/\delta n = 0$. Taking
this variational derivative of $E[n]$, we find
\begin{equation}
	\frac{\delta E}{\delta n} = \epsilon_K[n(\vr)] 
	+ \frac{e^2}{2\kappa} \int d\vr' \frac{n(\vr')}{|\vr-\vr'|}
	+ \frac{e^2}{2\kappa} V_D(\vr) - \mu.
\end{equation}

\begin{figure}
	\includegraphics[width=0.48\columnwidth]{%
		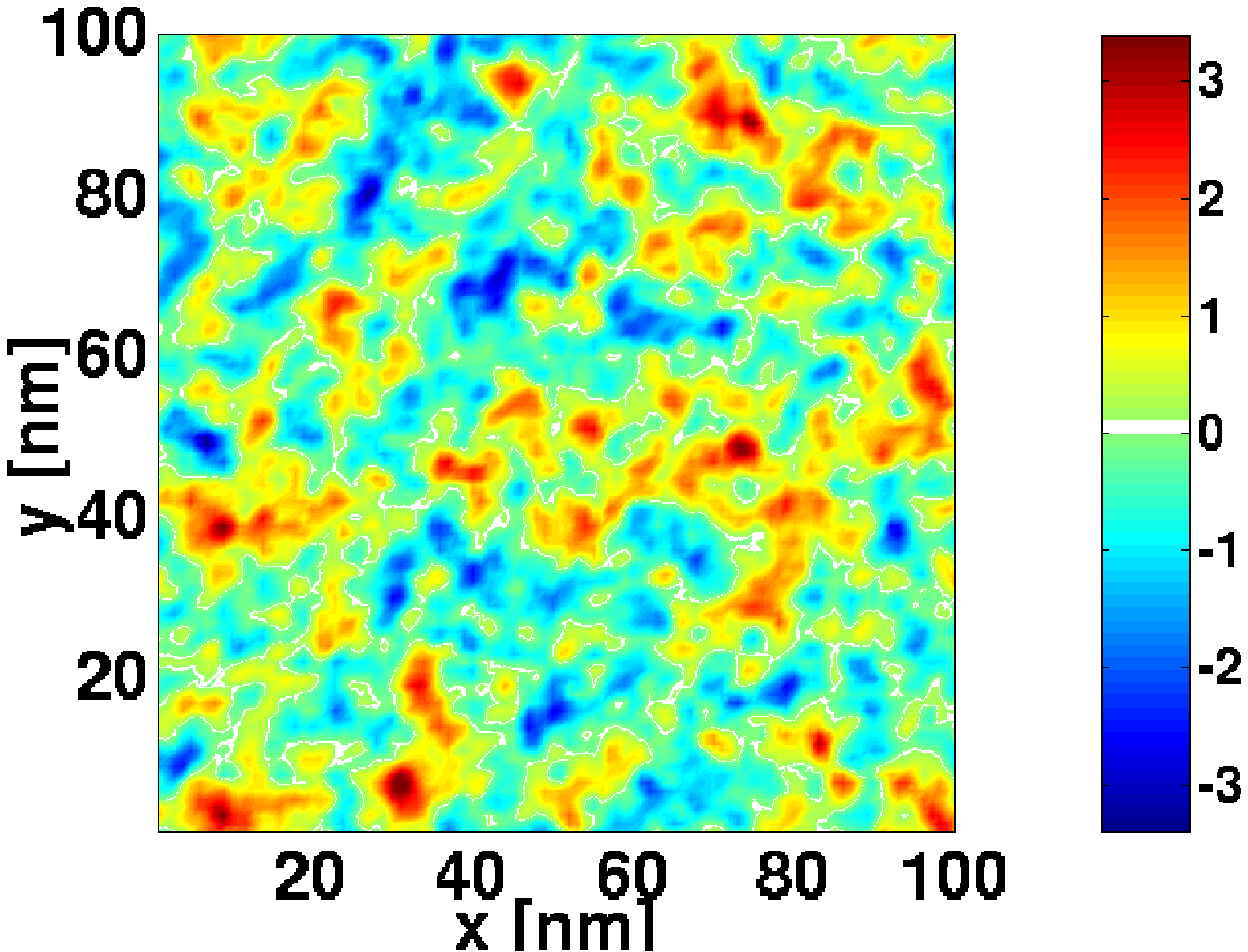}\hfill%
	\includegraphics[width=0.48\columnwidth]{%
		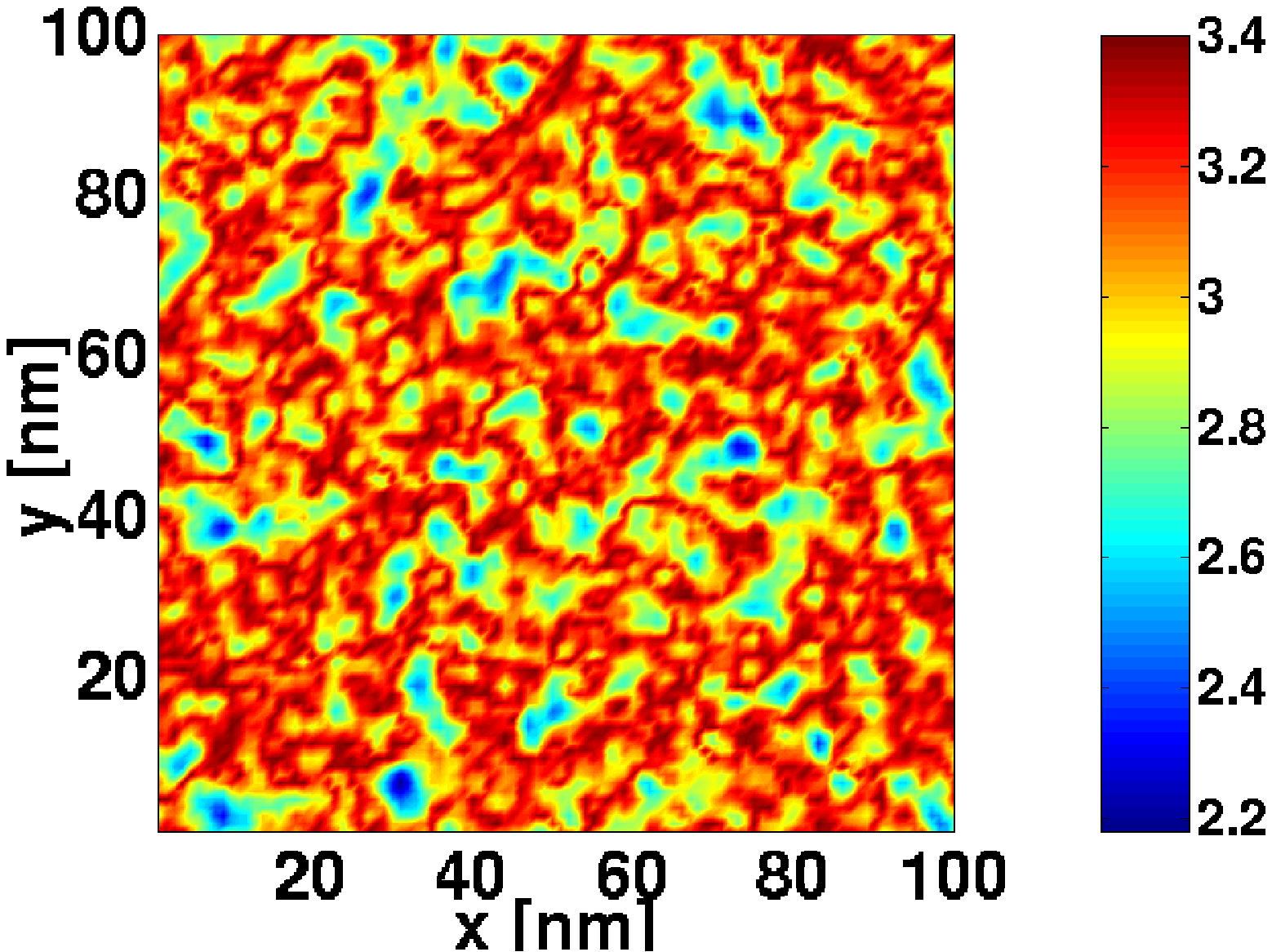}\\%
	\includegraphics[width=0.48\columnwidth]{%
		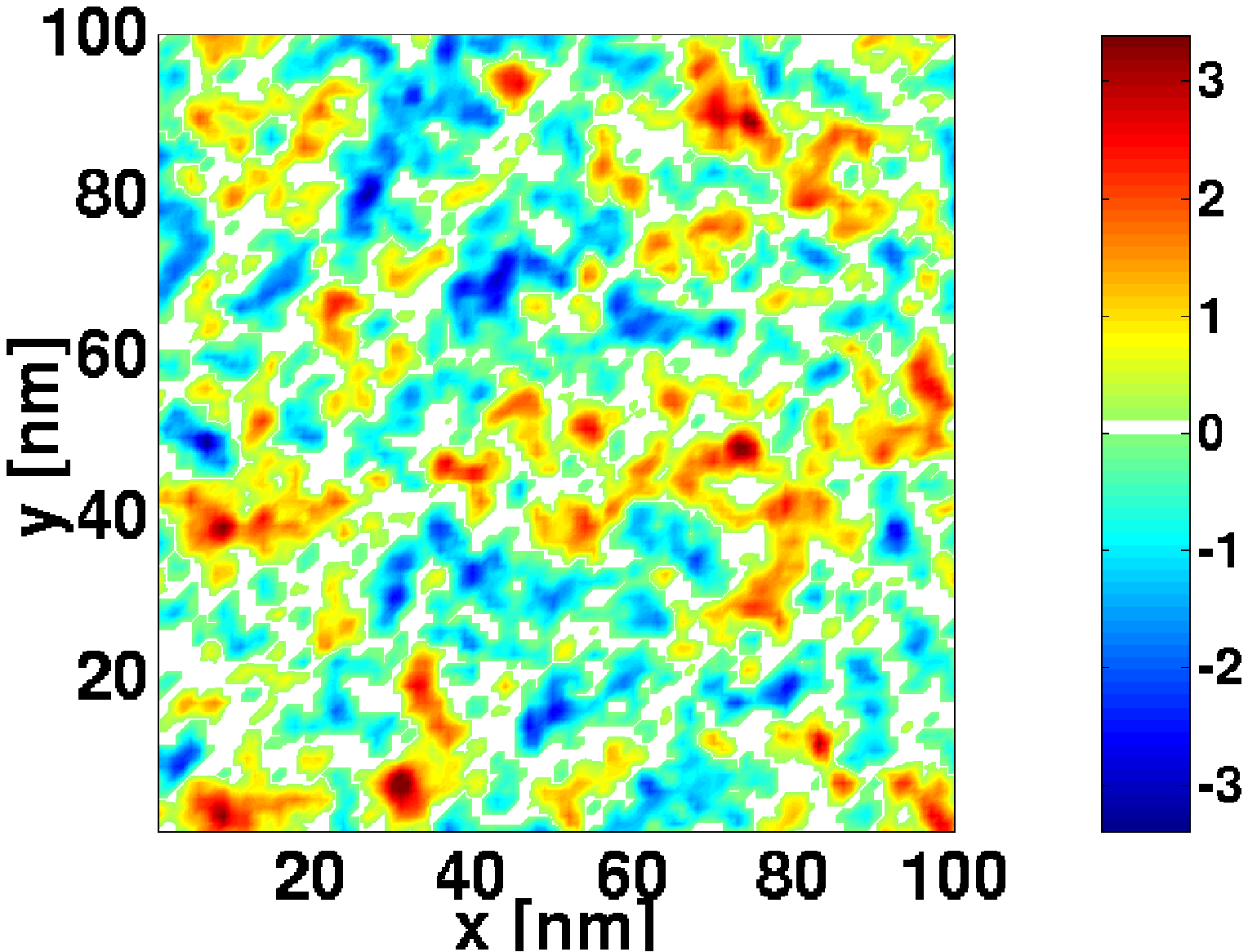}\hfill%
	\includegraphics[width=0.48\columnwidth]{%
		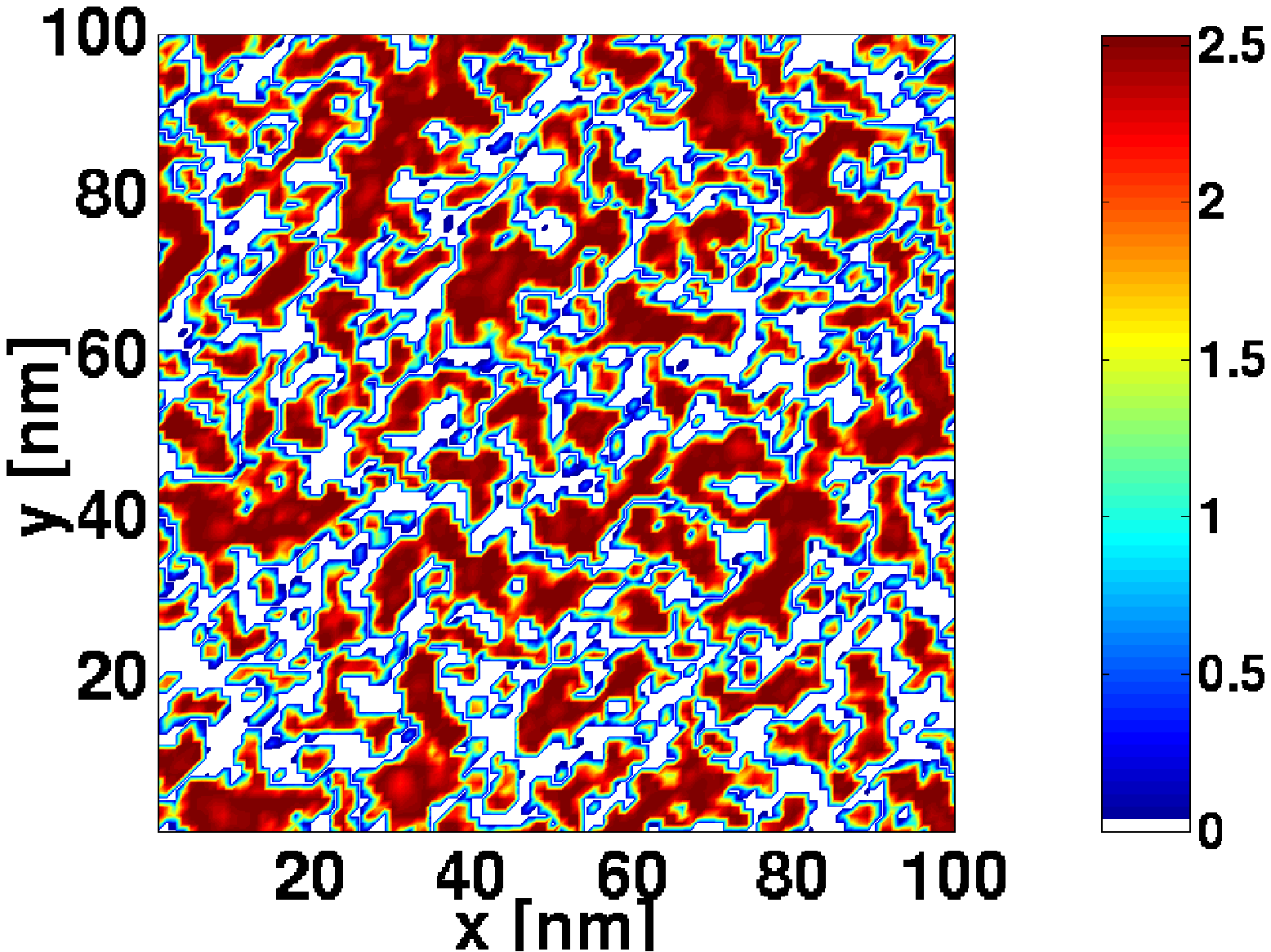}\\%
	\includegraphics[width=0.48\columnwidth]{%
		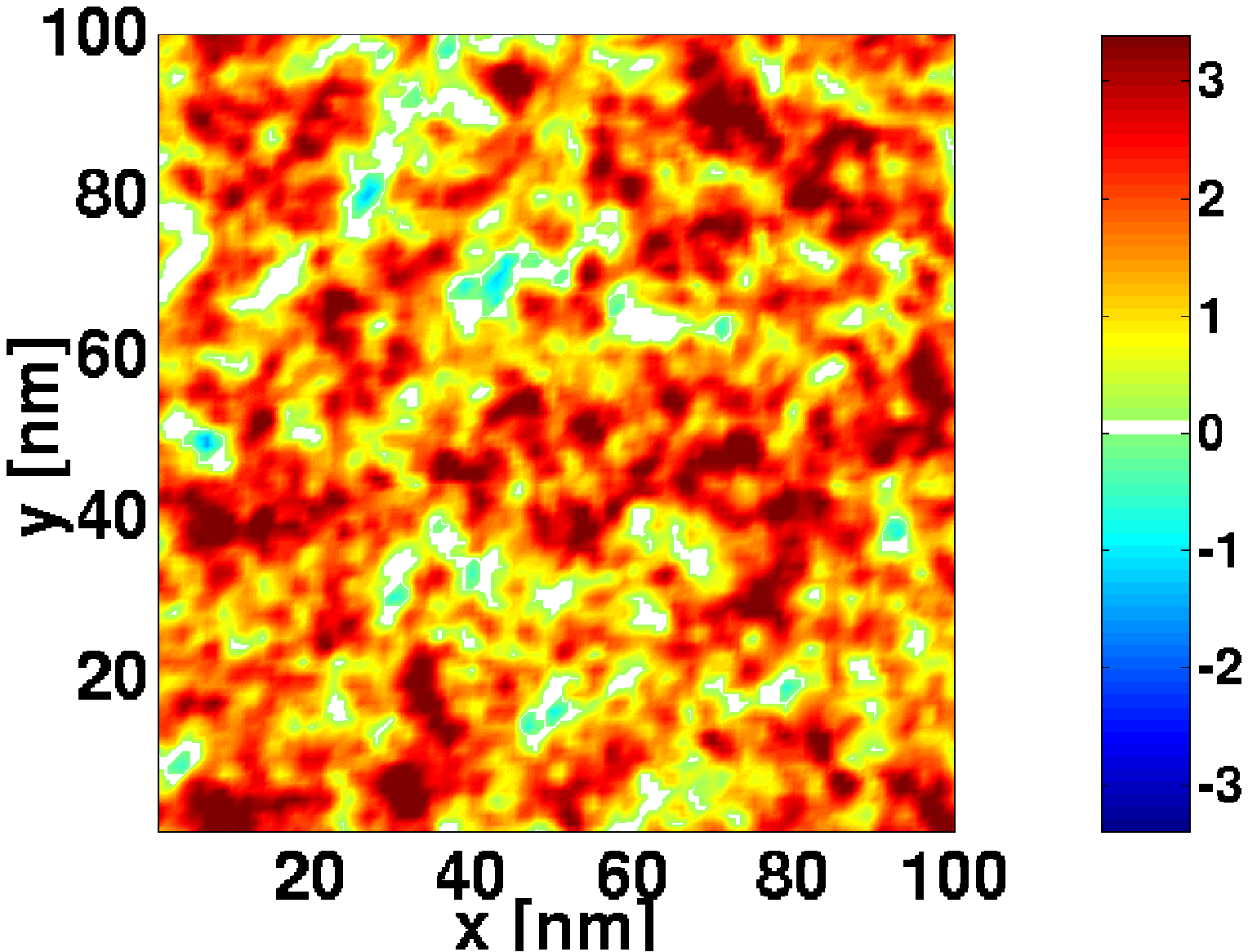}%
	\hfill%
	\includegraphics[width=0.48\columnwidth]{%
		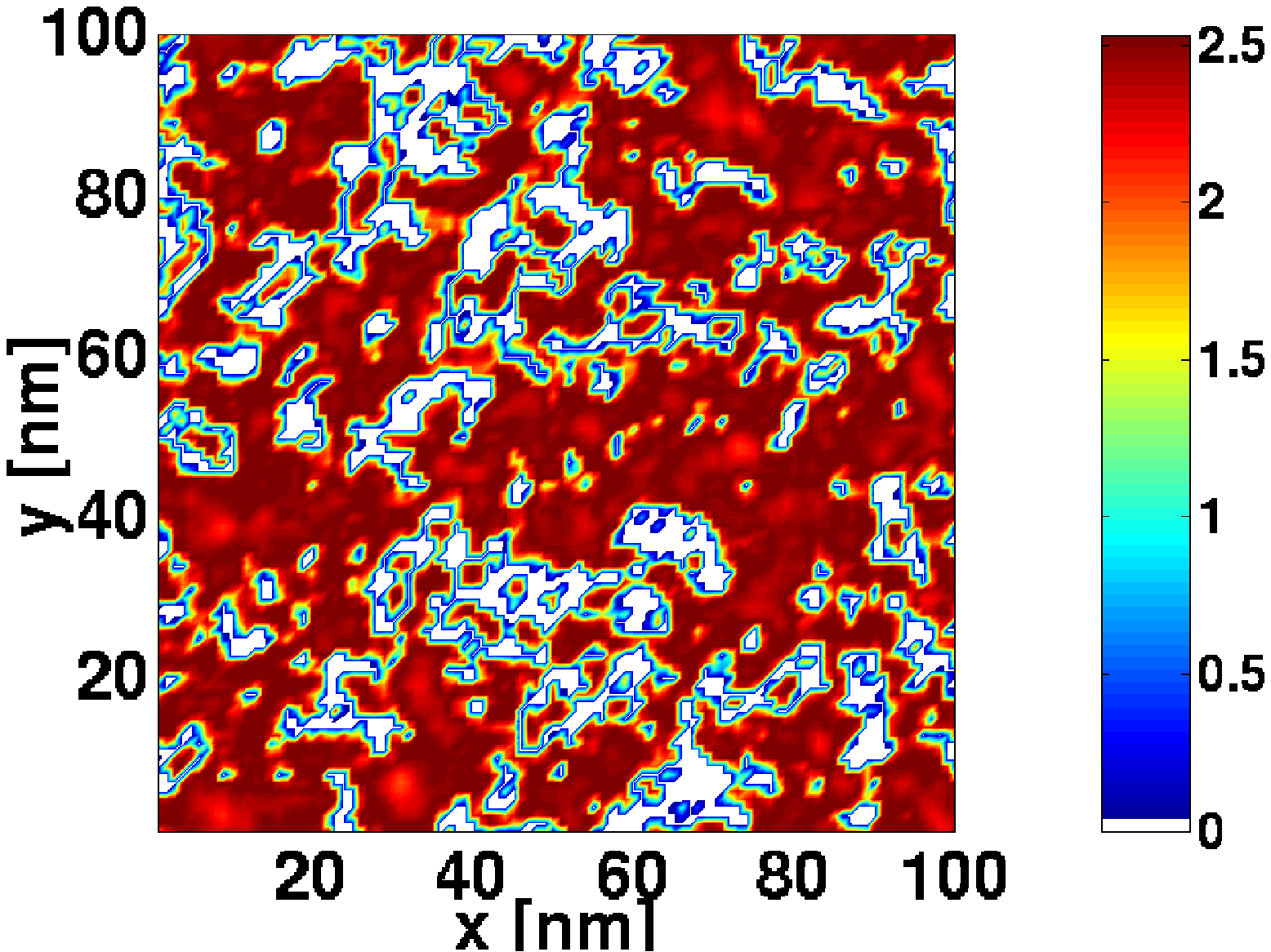}\\%
	\includegraphics[width=0.48\columnwidth]{%
		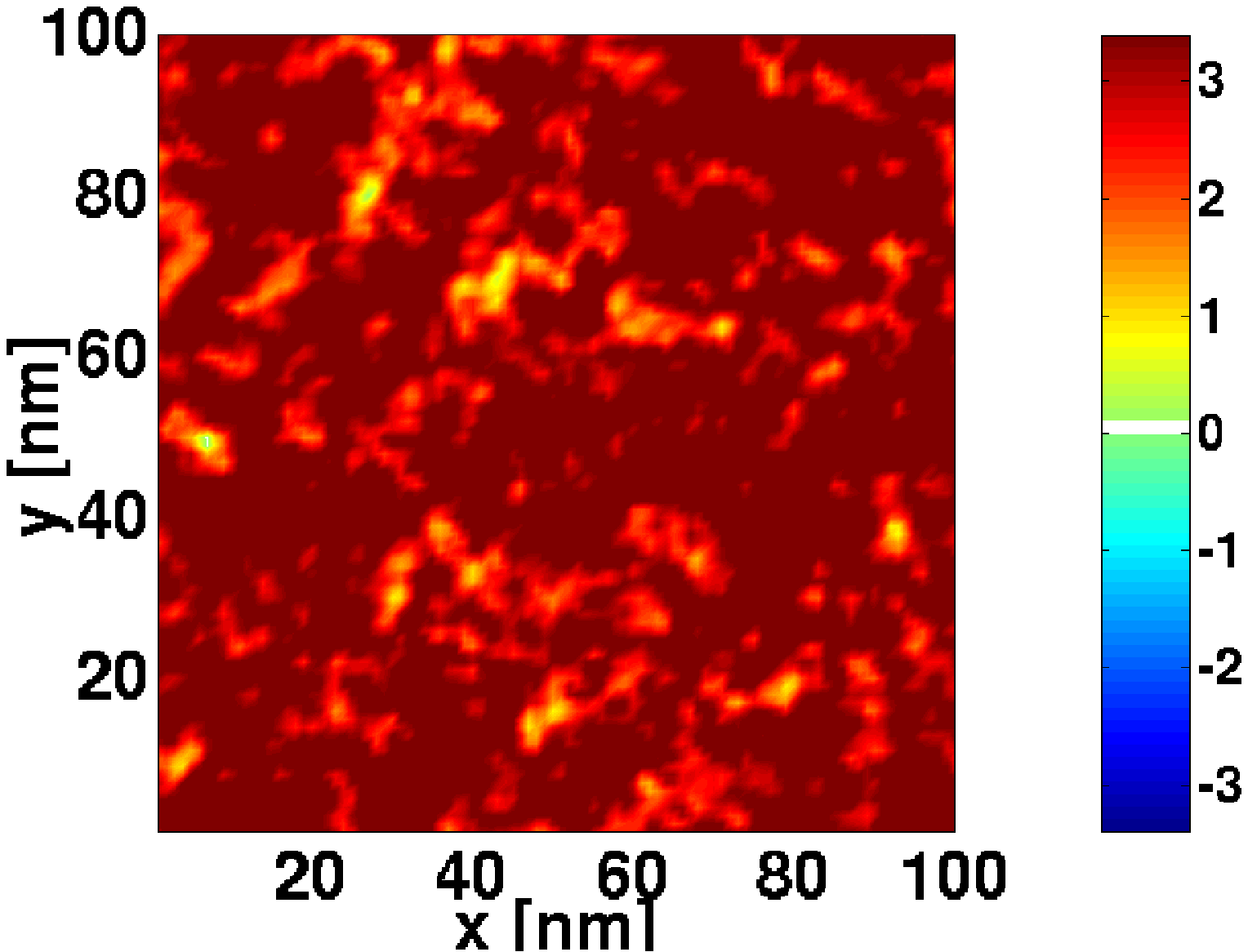}%
	\hfill%
	\includegraphics[width=0.48\columnwidth]{%
		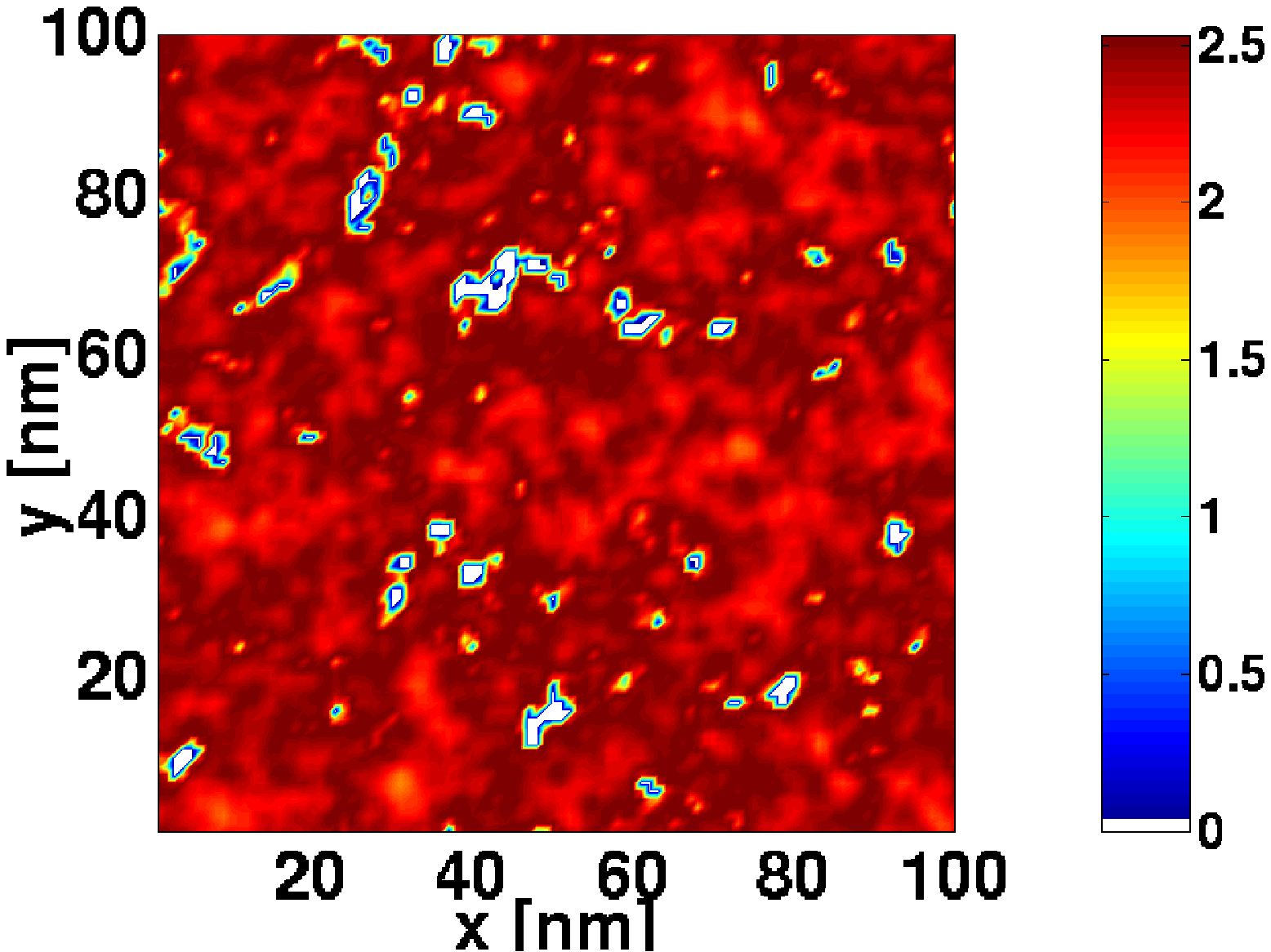}
	\caption{(Color online) (Left column) Spatial density profile, and
	(Right column) spatial $\dmudn$ profile for a single disorder
	realization. The units of the color bars are $10^{12}\cmsq$ for the
	density and $\eV\,\mathrm{nm}^2$ for $\dmudn$, and $\nimp=3\times
	10^{11}\cmsq$ throughout.
	The incompressible regions where $n=0$ are shown in white. 
	The first row has $u=0$ and $\nav=0$; 
	the second row has $u=40\meV$ and $\nav=0$;
	the third row has $u=40\meV$ and $\nav\approx 2\times 10^{12}\cmsq$;
	and the fourth row has $u=40\meV$ and $\nav\approx 4\times
	10^{12}\cmsq$.
	\label{fig:spatial}}
\end{figure}
Within our formalism it is fairly easy to assume the presence of spatial
correlations among the impurities, and the presence of correlations has
important effects on the transport properties of BLG; \cite{li-prl107,
yan-prl107, li-ssc152} however they do not modify the qualitative
effects that the disorder has on the compressibility.

Assuming that the clean $\dmudn$ is valid locally, in Fig.
\ref{fig:spatial} we show the spatial profile of the density
distribution (left column) and $\dmudn$ (right column) for a single
realization of disorder with $\nimp = 3\times 10^{11}\cmsq$ for
the gapless regime (first row) and the gapped case with
$u=40\meV$ and three values of the global charge density. 
The white regions are the parts of the graphene where the local density
is zero and hence the graphene is incompressible. It is immediately
noticeable that in the presence of a band gap (second row) there are
large incompressible regions which are not present in the gapless case
(even at zero excess carrier density),
and these regions persist even when the average charge density $\nav$ is
significant ($\nav \approx 2\times 10^{12}\cmsq$, third row) and are
still just visible when $\nav \approx 4\times 10^{12}\cmsq$ (fourth
row). 
More accurate functional methods than the TFT will not give
substantially different values for the ratio of the sample that is
covered by insulating (incompressible) regions, which we shall show is
the dominant factor in determining $\dmudn$ at low density. 
All these methods can do is to give slightly different values of $n$ and
density profiles inside metallic regions, both of which have very little
influence on $\dmudn$ for the situation of interest.

\begin{figure}
	\includegraphics[]{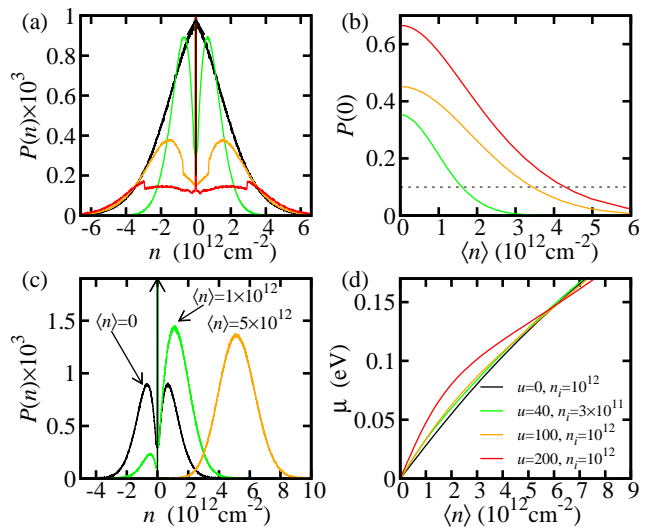}
	\caption{(Color online.) (a) $P(n)$ at the charge neutrality point
	$\nav=0$ for different values of $u$ and $\nimp$. For $u\neq 0$, 
	the arrow at $n=0$ represents the very narrow peak whose height is
	orders of magnitude larger than the $y$-scale used.
	(b) $P(n=0)$ for parameters corresponding to Fig.
	\ref{fig:perttheo}(b) to \ref{fig:perttheo}(d). A dotted line at
	$P(0)=0.1$ provided as a guide to the eye.
	(c) Evolution of $P(n)$ with $\nav$ for $u=40\meV$ and $\nimp =
	3\times 10^{11}\cmsq$.
	(d) $\mu$ as a function of doping. The legend in (d) also applies to
	panels (a) and (c).
	\label{fig:pplots}}
\end{figure}

By considering many disorder realizations we can calculate disorder
averaged quantities and, in particular, the probability distribution
function of the local carrier density $P(n)$. This function can then be
used to compute the average density $\nav = \int n'P(n')dn'$. As shown
in Fig. \ref{fig:pplots}(a), $P(n)$ is trimodal for $\nav=0$ in the
gapped regime: it exhibits a large peak shown by an arrow at $n=0$ that
quantifies the fraction of the sample ocupied by insulating regions and
two smaller and broader peaks centered around values of $n$ which are
determined by the nonlinear screening of the disorder potential and
which therefore depend on both $u$ and $\nimp$. Figure
\ref{fig:pplots}(a) also shows that $P(n)$ has a jump for $n=\tilde n$.
As the doping increases, the fraction of the sample area covered by
insulating regions $P(n=0)$ decreases as shown in Fig.
\ref{fig:pplots}(b). Notice the factor of $10^3$ difference in the
vertical scale between Figs. \ref{fig:pplots}(a) and
\ref{fig:pplots}(b).  
The evolution of $P(n)$ with $\nav$ is shown in Fig.
\ref{fig:pplots}(c). For finite doping, the distribution becomes
asymmetric in $n$ and becomes unimodal only at relatively large doping.
In the unimodal regime, $P(n)$ is approximated closely by a Gaussian
centered around $\nav$.

Experimental probes such as the capacitance measurements and scanning
SET microscopy, simultaneously probe an area of the sample which is
significantly larger than the puddles size as predicted in the TFT and
measured by STM. \cite{zhang-y-natphys5, deshpande-prb79, xue-natmat10,
deshpande-apl95}
Therefore, an averaging procedure must be implemented to replicate
$\dmudn$ as a function of
$\nav$ as measured in those experiments. By disorder averaging the TFT
results, we obtain the dependence of the average chemical potential
$\muav$ (which is identical to $\mu$ in Eq. \eqref{eq:Efunc}) with
respect to the average density $\nav$. Because of the non-linear
screening, the relation between $\muav$ and $\nav$ is also modified by
the value of the gap and the strength of the disorder, as shown in Fig.
\ref{fig:pplots}(d). Thus, the TFT results clearly show the inhomogeneous
nature of the carrier density landscape in BLG in the presence of
disorder. Using the TFT we calculate the average $\left\langle \dmudn
\right\rangle = \frac{d\muav}{d\nav}$ which closely simulates the way in
which $\langle d\mu/dn \rangle$ is obtained in both capacitance
measurements \cite{henriksen-prb82, young-arXiv1004, young-prb84} and in
SET spectroscopy. \cite{martin-natphys4, martin-prl105}

\begin{figure}
	\includegraphics[]{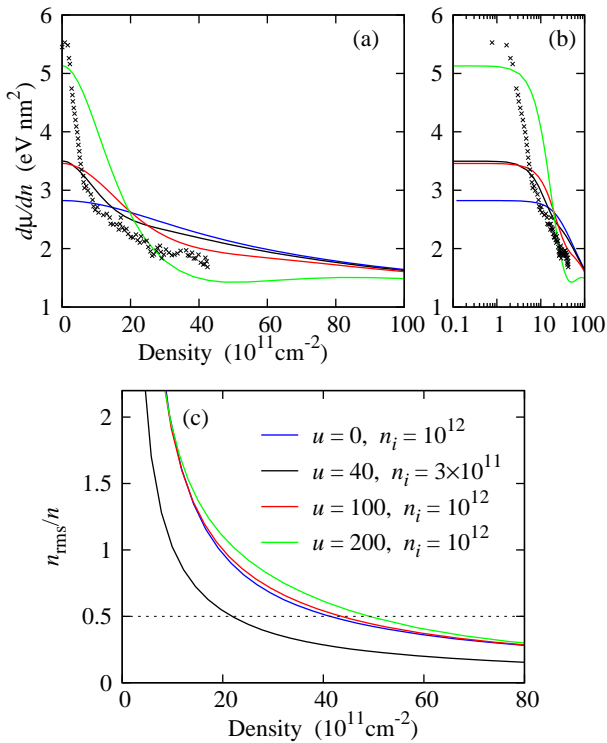}
	\caption{(Color online.) (a) The predicted $\dmudn$ in the TFT.
	(b) As in (a), but on a logarithmic scale to emphasize the
	low-density regime.
	(c) The density fluctuation due to disorder for the same parameters
	as in (a). 
	The dotted line at $n_{\nrms}/n = 0.5$ is included as a guide
	to the eye.
	\label{fig:tft}}
\end{figure}

Figure \ref{fig:tft}(a) shows the calculated $\dmudn$ using the TFT for
the same four sets of parameters as in Fig. \ref{fig:perttheo}. We
immediately see that it exhibits qualitatively different behavior from
the BA and SCBA, and at low density it shows a broad peak in qualitative
agreement with the experimental data. 
We stress that (as mentioned above) the data in Ref.
\onlinecite{henriksen-prb82} is the only appropriate data for direct
comparison with our theories, but since various parameters of the
experimental system are unknown, we cannot expect quantitative agreement
between our calculations and the measured data.
Comparison of Fig. \ref{fig:perttheo} with Figs. \ref{fig:tft}(a) and
\ref{fig:pplots}(b) shows that the deviation of the perturbation
theory from the TFT occurs when $P(0) > 0.1$, indicating that the
presence of insulating,
incompressible regions is the dominating feature of the compressibility
at low density. In Fig. \ref{fig:tft}(c) we show our TFT calculated
density fluctuation characterized by the root-mean-square value $\nrms$
as a function of density.
This clearly establishes that, when a band gap is present and as the
fluctuations (i.e. inhomogeneity) become large with decreasing average
density, the calculated TFT results for $\dmudn$ start deviating
substantially from the many-body perturbative ensemble averaged results,
and when $\nrms/n \approx 0.5$, one must carry out the nonlinear
screening theory to obtain the qualitatively correct features for the
compressibility.

\section{Discussion \label{sec:discussion}}

Therefore, we identify two different reasons for the failure of the
standard diagrammatic methods in gapped inhomogeneous systems. The first
is the presence of strong inhomogeneity characterized by the parameter
$\nrms/n$. When $\nrms/n > a$ with $a\sim 1$, the ground state cannot be
assumed to be homogeneous and therefore implicit translational
invariance incorporated in the disorder averaging step of the
diagrammatic theory fails to qualitatively describe the experimental
situation. The exact value of $a$ depends on the experimental quantity
under consideration and the details of the experimental conditions. The
second reason for failure is the existence of a random mixed
inhomogeneous state where insulating (incompressible) and metallic
(compressible) regions coexist in the presence of a band gap of the order
of or smaller than the disorder strength. The diagrammatic methods fail
because they cannot account for this mixed state. For the
compressibility, comparison of Fig. \ref{fig:pplots}(b) with Fig.
\ref{fig:perttheo} shows that the critical fraction of the sample area
to be covered by insulating regions for the perturbative theories to give
strong qualitative disagreement with the experiments is $P(0)>0.1$.
Thus, disorder has a much stronger qualitative effect at low carrier
densities for gapped bilayer graphene \cite{henriksen-prb82,
martin-prl105} than for the monolayer \cite{martin-natphys4} since BLG, by
virtue of being a gapped system, can be in the random mixed state which
is not accessible for ungapped systems.

In conclusion, we have demonstrated that the nonlinear nature of the
screening of an external disorder potential in gapped bilayer graphene
and the resulting charge inhomogeneity are crucial in understanding the
ground state electronic properties for a wide range of experimentally
relevant carrier densities. In particular, standard many-body
diagrammatic techniques assume that the density profile is homogeneous
in both the screening and the Green's function, and therefore give
qualitatively incorrect predictions for the compressibility in the
presence of an external band gap. In contrast, the TFT retains the
inhomogeneity and non-linear screening of the density distribution in
the energy functional $E[n(\vr)]$ and therefore captures the essential
physics of the system.
We emphasize that this particular averaging procedure discussed in Sec.
\ref{sec:tft}, simulating the experimental conditions, is simply
inaccessible to any type of theories invoking a homogeneous charge
landscape to obtain many-body self-energy or broadening.
We also point out that although we have presented results where the
disorder potential is induced by random charged impurities, our general
conclusion will remain valid for any form of disorder which produces a
scalar potential perturbation to the clean Hamiltonian, such as
corrugations in the graphene sheet. \cite{gibertini-prb85}
Finally, although we have described calculation of the compressibility
(and equivalently $\dmudn$) the general logic of our argument applies
to other observable quantities also. For instance, application of the
Kubo formula for transport with Green's functions derived in the same
way as in Sec. \ref{sec:perttheo} will suffer from similar problems in
the inhomogeneous regime. 
In this example, an effective medium theory \cite{rossi-prb79} or a full
quantum transport analysis that explicitly takes into account the
inhomogeneities \cite{rossi-prl109} should be used instead.

\begin{acknowledgments}
We thank US-ONR and NRI-SWAN for financial support. E.R. acknowledges
support from the Jeffress Memorial Trust, Grant No. J-1033. E.R. and DSLA
acknowledge the KITP, supported in part by the National Science
Foundation under Grant No. PHY11-25915, where part of the work was
carried out. Computations were conducted in part on the SciClone Cluster
at the College of William and Mary.
\end{acknowledgments}

\bibliography{bibtex-sorted}

\end{document}